\def\papertitle{Object-based synthesis of scraping and rolling sounds based on non-linear physical constraints }
\def\paperauthorA{Vinayak Agarwal}
\def\paperauthorB{Maddie Cusimano}
\def\paperauthorC{James Traer}
\def\paperauthorD{Josh McDermott}
\documentclass[twoside,a4paper]{article}
\usepackage{etoolbox}
\usepackage{dafx_20}
\usepackage{amsmath,amssymb,amsfonts,amsthm}
\usepackage{euscript}
\usepackage[T1]{fontenc}
\usepackage[utf8]{inputenc}
\usepackage{nimbusserif}
\usepackage{ifpdf}
\usepackage[english]{babel}
\usepackage{caption}
\usepackage{subfig} % or can use subcaption package
\usepackage{color}
\usepackage{float}
\usepackage{mathtools}
\usepackage{textgreek}

\input glyphtounicode
\ninept

\newcounter{numauth}
\newcounter{listcnt}
\newcommand\authcnt[1]{\ifdefined#1 \stepcounter{numauth} \fi}

\newcommand\addauth[1]{
\ifdefined#1 
\stepcounter{listcnt}
\ifnum \value{listcnt}<\value{numauth}
\appto\authorslist{, #1}
\else
\appto\authorslist{~and~#1}
\fi
\fi}
\authcnt{\paperauthorB}
\authcnt{\paperauthorC}
\authcnt{\paperauthorD}
\authcnt{\paperauthorE}
\authcnt{\paperauthorF}
\authcnt{\paperauthorG}
\authcnt{\paperauthorH}
\authcnt{\paperauthorI}
\authcnt{\paperauthorJ}
\def\authorslist{\paperauthorA}
\addauth{\paperauthorB}
\addauth{\paperauthorC}
\addauth{\paperauthorD}
\addauth{\paperauthorE}
\addauth{\paperauthorF}
\addauth{\paperauthorG}
\addauth{\paperauthorH}
\addauth{\paperauthorI}
\addauth{\paperauthorJ}
\usepackage{times}
\newif\ifpdf
\ifx\pdfoutput\relax
\else
   \ifcase\pdfoutput
      \pdffalse
   \else
      \pdftrue
\fi

\ifpdf % compiling with pdflatex
  \usepackage[pdftex,
    pdftitle={\papertitle},
    pdfauthor={\authorslist},
    pdfsubject={Proceedings of the 23rd International Conference on Digital Audio Effects (DAFx-20)},
    colorlinks=false, % links are activated as color boxes instead of color text
    bookmarksnumbered, % use section numbers with bookmarks
    pdfstartview=XYZ % start with zoom=100% instead of full screen; especially useful if working with a big screen :-)
  ]{hyperref}
  \usepackage[pdftex]{graphicx}
\else % compiling with latex
  \usepackage[dvips]{epsfig,graphicx}
  \usepackage[dvips,
    pdftitle={\papertitle},
    pdfauthor={\authorslist},
    pdfsubject={Proceedings of the 23rd International Conference on Digital Audio Effects (DAFx-20)},
    colorlinks=false, % no color links
    bookmarksnumbered, % use section numbers with bookmarks
    pdfstartview=XYZ % start with zoom=100% instead of full screen
  ]{hyperref}
\fi
\usepackage[hypcap=true]{caption}
\title{\papertitle }

\fouraffiliations{
\paperauthorA \,\sthanks{Corresponding Author}}
{\href{https://meche.mit.edu}{Department of Mechanical Engineering} \\ Massachusetts Institute of Technology\\ Cambridge, US\\ {\tt \href{mailto:vinayaka@mit.edu}{vinayaka@mit.edu}}
}
{\paperauthorB \,}
{\href{https://bcs.mit.edu}{Department of Brain and Cognitive Sciences} \\ Massachusetts Institute of Technology\\ Cambridge, US\\ {\tt \href{mailto:mcusi@mit.edu}{mcusi@mit.edu}}
}
{\paperauthorC \,}
{\href{https://psychology.uiowa.edu/}{Department of Psychological and Brain Sciences} \\ University of Iowa\\ Iowa City, US\\ {\tt \href{mailto:james-traer@uiowa.edu}{james-traer@uiowa.edu}}
}
{\paperauthorD \,}
{\href{https://bcs.mit.edu}{Department of Brain and Cognitive Sciences} \\ Massachusetts Institute of Technology\\ Cambridge, US\\ {\tt \href{mailto:jhm@mit.edu}{jhm@mit.edu}}
}
\begin{document}
% more pdf-tex settings:
\ifpdf % used graphic file format for pdflatex
  \DeclareGraphicsExtensions{.png,.jpg,.pdf}
\else  % used graphic file format for latex
  \DeclareGraphicsExtensions{.eps}
\fi

%\makeatletter
%\pdfbookmark[0]{\@pdftitle}{title}
%\makeatother

\maketitle

\begin{abstract}
Sustained contact interactions like scraping and rolling produce a wide variety of sounds. Previous studies have explored ways to synthesize these sounds efficiently and intuitively but could not fully mimic the rich structure of real instances of these sounds. We present a novel source-filter model for realistic synthesis of scraping and rolling sounds with physically and perceptually relevant controllable parameters constrained by principles of mechanics. Key features of our model include non-linearities to constrain the contact force, naturalistic normal force variation for different motions, and a method for morphing impulse responses within a material to achieve location-dependence. Perceptual experiments show that the presented model is able to synthesize realistic scraping and rolling sounds while conveying physical information similar to that in recorded sounds.  
\end{abstract}

%Synthesizing these sounds is possible using detailed simulations of the motion, vibration and sound radiation of these objects. But perceptually compelling synthesis of these sounds may not need such detailed physical simulations.

%\section{Citations to use}
%
%James suggests (maybe) citing these somewhere (but otherwise approves the references as comprehensive):
%-- ren2013example - they anticipated our modal IR synthesis from measurements -> Vin isn't using sum of damped sinsusoids - instead he's using sinusoidal modelling and replicating the (arbitrarily evolving) mode powers from the recording 
%-- conan2012perceptual - they anticipated our perceptual experiments on synthetic contact sounds --> already covered by conan2014 (they have a lot of papers on the same method)
%-- reiss2019real - This is a short simple paper, but does have a non-trivial scraping sound synthesis algorithm in it. And Reiss and many of his students will likely be there. - put in introduction

\section{Introduction}
\label{sec:intro}

Collisions, scraping, and rolling are commonplace in daily life, and the sounds they produce convey information about physical events in the world. Often this information is uniquely available via sound. For instance, the visual evidence that an object is in contact with another is often ambiguous, but objects in contact will make sound if they move. 

The synthesis of contact sounds is accordingly important for a wide range of applications including virtual reality, game engines, auditory displays, and the training of machine perception systems. Such synthesis requires mapping physical variables (object materials, shapes, and motions) to sound. These generative models of sound could also contribute to theories of perception, which involves inferring physical causes from sound \cite{gaver1993world}, potentially by inverting internal generative models \cite{kersten2003inference}. 

Contact sound synthesis is possible via detailed but expensive physical simulations \cite{cadoz1993cordis,o2002synthesizing,zheng2011toward,bilbao2009numerical,james2006precomputed,manocha2009interactive,raghuvanshi2006interactive}. However, many applications require more efficient synthesis, as might be enabled by lower-dimensional characterizations of objects. Previous methods have either used signal-based heuristics \cite{van2001foley,conan2014intuitive} or physics-inspired intuitive modelling \cite{serafin2004sound,stoelinga2007psychomechanical,avanzini2005interactive,reiss2019real}. These models are able to convey some intended physical parameters, but thus far have not been fully perceptually convincing. 

\begin{sloppypar} We previously introduced a method to synthesize impact sounds by characterizing material-specific impulse response distributions, sampling from them, and convolving them with simple spring-based models of impact forces \cite{traer2019contact}. In that paper, we extended the method to synthesize scraping sounds using force derived from a surface depth map, but the sounds produced were not fully realistic, sounding excessively rough and not producing a fully accurate sense of motion.
\end{sloppypar}
Here we introduce a new method to synthesize scraping and rolling sounds from object properties. The key innovations are a nonlinearity in the calculation of the contact force from the surface depth map and the use of normal forces that more accurately reflect typical scraping motions. These are combined with a method for generating more realistic location-dependent impulse responses via morphing within the same material, similar to that proposed in \cite{pruvost2015perception}, and a periodic force produced by rolling objects due to misalignment of the center-of-mass with the geometric center \cite{rath2005continuous,van2001foley}. We term the approach `object-based', as it relies on perceptually-relevant macroscopic properties of objects and their motions.

The resulting sounds are substantially more realistic and recognizable than those from previous methods. We first describe and evaluate our scraping synthesis. We then show that the scraping synthesis can be extended to generate realistic rolling sounds. You can listen to demos of sounds from all our experiments at \url{http://mcdermottlab.mit.edu/scraping_rolling.html}.

\begin{figure*}[ht]
\center
\includegraphics[width=4.6in]{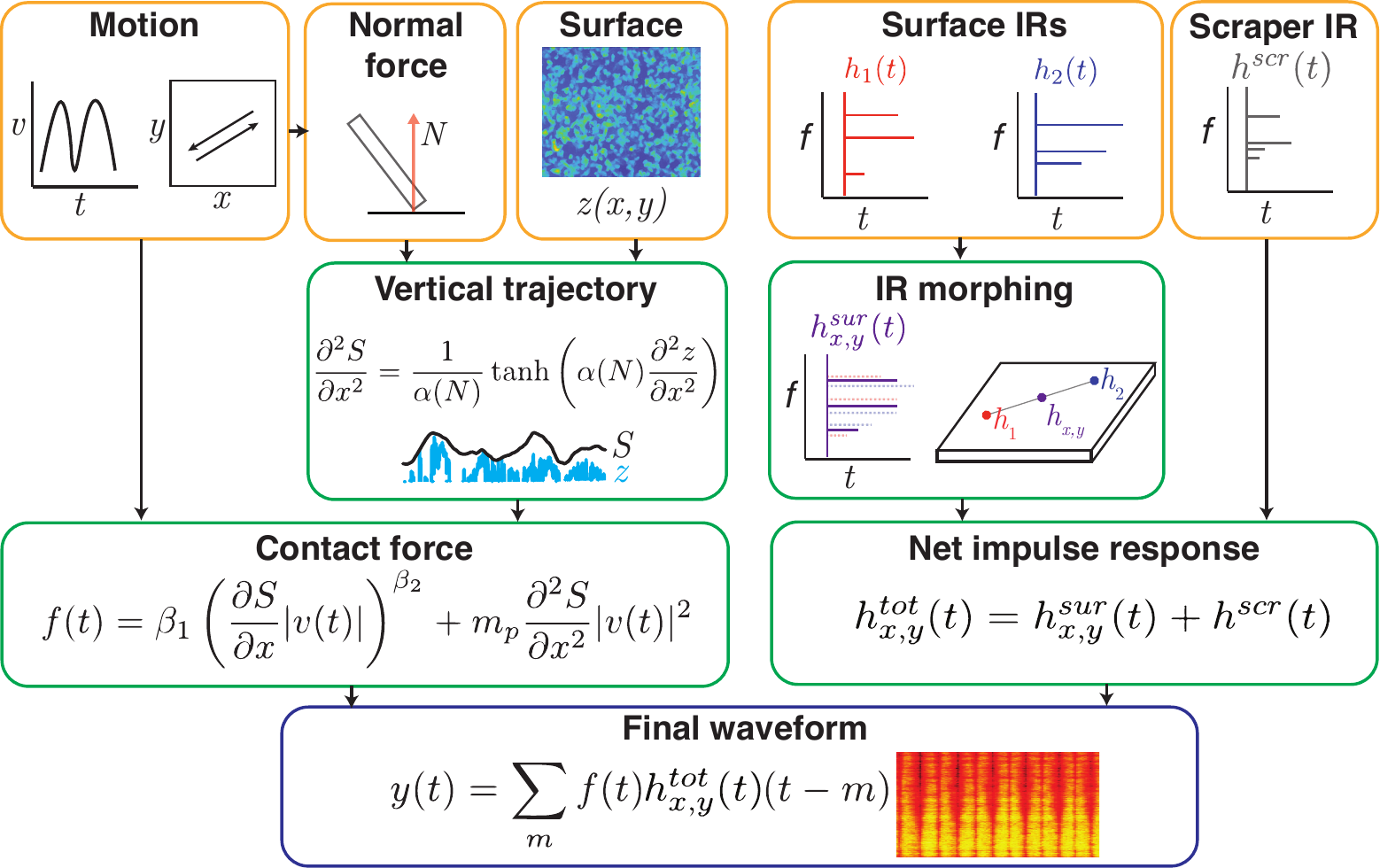}
\caption{\label{fig:scrap_mod}{\it Complete scraping synthesis model. Yellow boxes: inputs to model. Green boxes: intermediate representations computed from inputs. Blue box: sound waveform computed from contact force and net impulse response.}}
\end{figure*}

\section{Scraping Sound Synthesis Model}
\label{section:scraping}

% \begin{figure*}[ht]
% \center
% \includegraphics[width=3in]{all.pdf}
% \caption{\label{scrap_mod}{\it Complete scraping synthesis model}}
% \end{figure*}
We use a source-filter approach depicted in Figure \ref{fig:scrap_mod}. Similar to \cite{traer2019contact,conan2014intuitive,ren2010synthesizing}, we generate a sound waveform by the convolution of a contact force function and a net impulse response (IR). The contact force $f(t)$ is obtained by modelling the microscopic trajectory of the scraper on the scraped surface. We use a weighted sum of the impulse responses of both the scraper and the scraped surface to get the net impulse response of the interaction. To reflect the variation in object resonances with position, the impulse response of the surface changes continuously over the motion. We describe the force, impulse response and audio waveform calculation in more detail below. %The results suggest that our model is able to synthesize realistic scraping sounds which convey rich information about the relative motion.

\subsection{Contact force for scraping}
The contact between scraper and surface causes the two bodies to undergo continuous micro-collisions. Previous authors have sampled the friction force as $1/f^\beta$ noise \cite{ren2010synthesizing,van2001foley} or from a distribution over concentrated impact events \cite{conan2014intuitive,lagrange2010analysisrigid,lee2010analysis}. In our model, the only major assumption is that the scraper follows the surface without bouncing. We also assume that the masses of the scraper and surface do not change over the course of the motion. 

The microscopic interaction force is modelled in two parts - a vertical force $f_{\text{v}}$ (from pressing the scraper down) and a horizontal force $f_{\text{h}}$ (from horizontal micro-collisions with surface asperities). We also calculate an additional macroscopic normal force (\S\ref{section:nforce}), which modulates the microscopic interaction force.

The vertical force $f_v$ is estimated as:
\begin{align}{\label{scrap_force}}
    f_{\text{v}}(t) &= m_{\text{p}}\Ddot{S}(x,y) \\
    &= m_{\text{p}}\left(\frac{\partial^2 S(x,y)}{\partial x^2}|v_x(t)|^2 + \frac{\partial^2 S(x,y)}{\partial y^2}|v_y(t)|^2 \right)
\end{align}
where $m_p$ is the mass of the scraper, $S(x,y)$ is the scraper's vertical trajectory over the surface, and $v_x(t)$ and $v_y(t)$ are the velocities of the scraper in the $x$ and $y$ directions respectively (which define the plane of the surface).

We assume the horizontal contact force $f_h$ to originate from the horizontal collisions of the scraper and steep asperities of the surface texture. Based on previous literature \cite{traer2019contact}, we assume the following relation:
\begin{align}
    f_h(t) = \beta_1 \left| v_x(t) \frac{\partial S(x,y)}{\partial x}+v_y(t) \frac{\partial S(x,y)}{\partial y}\right| ^{\beta_2}
\end{align}
where the partial derivatives with respect to the position ($x$,$y$) reflect the slopes of the surface texture and determine the intensity of the micro-impacts. The velocities $v_x$ and $v_y$ determine the density of these impacts in time. $\beta_1$, $\beta_2$ are free parameters that we set to 0.05 and 1, respectively. 

We model the contribution of the horizontal and vertical force components to the resulting sound as a convolution with the same impulse response. Because the convolution operator is linear and the resulting sound waveforms add together ($y(t) = f_{\text{v}}(t) * h(t) + f_{\text{h}}(t) * h(t)$), for notational convenience we denote a total force term given by a scalar sum $f_{\text{tot}}(t) = f_{\text{v}} + f_{\text{h}}$.  

\subsection{Trajectory of the scraping object}
\label{section:nonlinearity}
%MC: Another place we need to be really clear with our language is the "vertical trajectory" of the probe near the surface, and the path that the probe takes over the surface - let' make sure we're consistent throughout

%also check for consistency in "scraper probe tip" vs "scraper" etc.

One of the most critical components of our model is the vertical trajectory $S(x,y)$ of the scraper near the surface. Most previous models have assumed the scraper's vertical trajectory to be the same as the surface depth profile $z(x,y)$ \cite{traer2019contact,conan2014intuitive}. In our model, we use $z(x,y)$ to derive the estimated vertical trajectory of the scraper but additionally incorporate nonlinear physical constraints on the scraper-surface interaction. As in previous work, we assume the scraper to be in contact with the surface at a single point.

We obtained surface depth profiles for several everyday materials using a scanning confocal microscope (Keyence VK-X260K, horizontal resolution of 5.6 \textmu m, vertical resolution of 0.1 nm) \cite{traer2019contact}.  %Since the rich surface textures of these materials can alter the spectral make-up of the final sound, we preferred using it over fractal noise. 
In the future, we plan to synthesize surface depth maps as textures using relevant texture statistics of measured depth profiles \cite{mcdermott2011sound}.

% \subsubsection{Limiting surface curvatures}
The assumption that the scraper exactly follows the surface depth profile $z$ is physically unrealistic for two reasons: 1) the forces involved would be exceptionally high near the locations on the surface where the slopes change rapidly, and 2) both the surface and scraper  materials would have to be abnormally elastic.

\begin{figure}[ht]
\centerline{\includegraphics[scale=0.45]{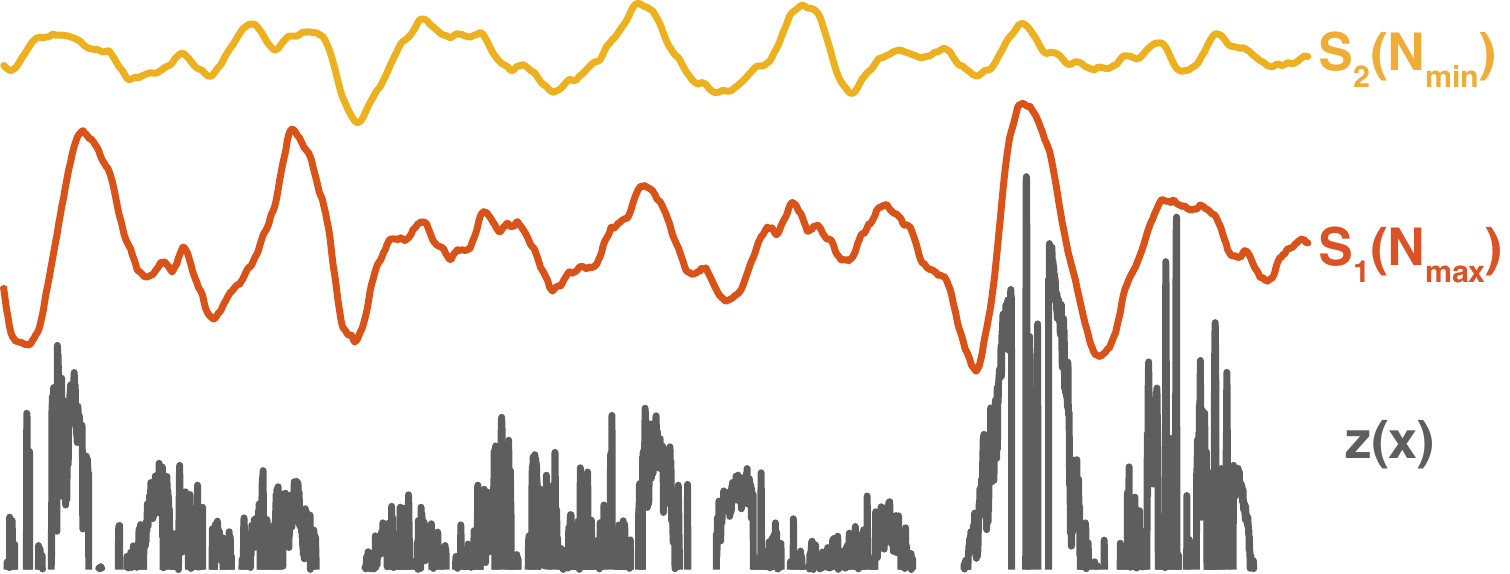}}
\caption{\label{fig:surf_nonlin}{\it Trajectory $S$ of the scraper point. The trajectory is determined by the surface depth profile ($z(x)$, shown in grey) and the normal force $N$. Larger normal forces (red) produce more extreme scraper trajectories than smaller normal forces (yellow).}}
\end{figure}

%MC: One small note is that we keep changing between "scraper tip", "probe tip" etc., we should just decide on one and keep it that way.

To solve these issues, we have to constrain the allowed curvatures of the scraper trajectory $S(x,y)$ to limits that are physically realizable given bounded applied force. The applied force bounds the curvature of the turns that the scraper can take over the surface asperities, limiting its acceleration and hence the second derivative of $S$ (Figure \ref{fig:surf_nonlin}). To achieve this, we used a tanh non-linearity with limits dictated by the elasticity of the material and the macroscopic normal force $N$.
\begin{align}
    \frac{\partial^2 S(x,y)}{\partial x^2} &= \frac{1}{\alpha_x(N)}\tanh\left(\alpha_x(N) \frac{\partial^2 z(x,y)}{\partial x^2}\right) \\
    \frac{\partial^2 S(x,y)}{\partial y^2} &= \frac{1}{\alpha_y (N)}\tanh\left(\alpha_y (N) \frac{\partial^2 z(x,y)}{\partial y^2}\right)
\end{align}
$\alpha_x$ and $\alpha_y$ change the limits of the non-linearity depending on the normal force:
\begin{align}
 \alpha(N) &= (1- \nu) \alpha_{\text{max}} + \nu \alpha_{\text{min}} \\
 \nu &= \left(\frac{N-N_{\text{min}}}{N_{\text{max}}-N_{\text{min}}}\right) ^\zeta
\end{align}

 $N_{\text{max}}$ and $N_{\text{min}}$ are determined by the variation in normal force described in the next section; for constant $N$ we substitute fixed values for $\alpha_x$ and $\alpha_y$. The current synthesis process is dependent on a judicious choice of $\zeta$, $\alpha_{\text{max}}$ and $\alpha_{\text{min}}$ (we found values of 0.95, 0.05, and 0.01, respectively, to yield realistic sounds). In the future, we aim to estimate these parameters from the applied vertical force and hardness of the material. 

This procedure is a form of soft clipping which introduces high-frequency components, presumably because the tanh function does not limit the higher-order derivatives of the scraper trajectory to an extent that is physically correct. To obtain realistic sounds, we found it necessary to apply a Gaussian moving average following the non-linearity to remove these high-frequency components. The window size in each dimension was proportional to $\alpha_x$ and $\alpha_y$, with an average half window size of 5 samples at a sampling rate of 44.1kHz. Alternative choices of non-linearity might obviate the need for this additional step.
%Note to James- please look over this!

\subsection{Effect of macroscopic normal force}
\label{section:nforce}

\begin{figure}[ht]
\centerline{\includegraphics[scale=1]{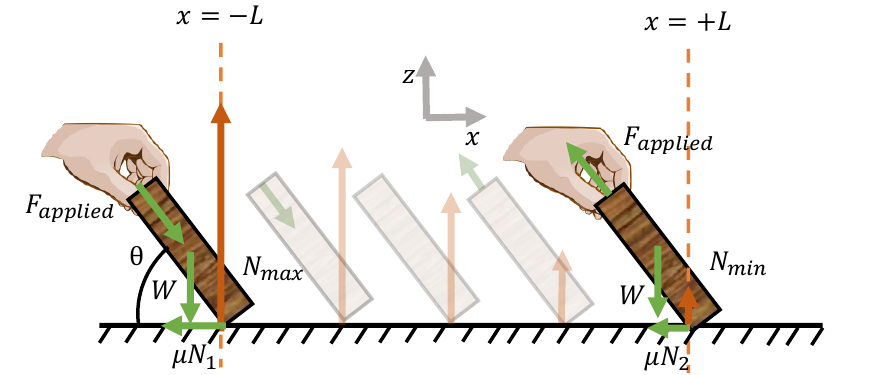}}
\caption{\label{fig:SHM_ex}{\it Macroscopic forces on the scraper when undergoing human-produced simple harmonic motion (as when scraping a held object back and forth on a surface). The normal force (red vector) varies with position due to the variation in the applied force $F_{\text{applied}}$.}}
\end{figure}

The applied force dictates how closely the scraper tracks the surface, which in our model is reflected in the force-dependent nonlinearity parameters $(\alpha_x, \alpha_y)$ that influence the modulation of the normal force (Figure \ref{fig:surf_nonlin}). The applied force can vary over the scraping trajectory, in part due to constraints on how the human hand applies forces. For example, consider back-and-forth scraping with a cylindrical object held by hand (Figure \ref{fig:SHM_ex}). This action can be approximated as simple harmonic motion with an angular frequency $\omega$. We calculate the resulting normal force assuming that 1) the applied force is along the length of the cylinder 2) the force is proportional to the horizontal acceleration of the cylinder, and 3) the cylinder is at an angle $\theta$ from the horizontal. As a result, the force is maximal when close to the torso and minimal when furthest away. The normal forces at both ends of the trajectory are then given by:  
\begin{align}
  N_{\text{max}} &= \frac{mg+\omega^2 mL \tan \theta}{1- \mu \tan{\theta}} \\
  N_{\text{min}} &= \frac{mg-\omega^2 mL \tan \theta}{1- \mu \tan{\theta}}
\end{align}
This normal variation is characteristic of back-and-forth scraping motion (e.g. from sanding or scrubbing). The normal force variation influences the scraping sound via $\alpha_x$ and $\alpha_y$ as described in the previous section. For each motion trajectory, we similarly modelled the normal force variation assuming a scraping cylinder held at a fixed angle to the surface.

% Vin - Calculated normal force for other trajectories based on the assumption that a human is performing the scraping motion with the probe at an acute angle with the surface

\subsection{Morphing impulse responses}

The impulse responses on the surface depend on the location on the surface currently being excited, because the boundary conditions change with the scraper position. Our measurements of object impulse responses indicated that the modal frequencies and their powers typically vary smoothly with location. To generate an impulse response for an arbitrary surface location, we thus morph from one measured impulse response to another by smoothly interpolating between both the modal frequencies and powers (similar to \cite{pruvost2015perception}), in contrast to our previous work where we cross-faded impulse responses \cite{traer2019contact}.

We used impulse responses measured in previous work \cite{traer2019contact} at different locations on the surfaces of objects. We then extracted the mode parameters of the 50 strongest modes using sinusoidal modelling, giving us the frequency $f_i$ of each mode and its time-dependent mode power $A_i(t)$. To obtain the mode parameters for an impulse response at a location intermediate between two measured impulse responses we logarithmically interpolated between the measured modes in both frequency and amplitude. The impulse response for the location was then synthesized as a superposition of decaying sinusoids using these morphed parameters.

As the scraper applies a force on the surface, the surface applies an equal and opposite reaction force on the scraper. If the scraper is not damped, it also contributes to the overall sound because of the effect of this forcing on the scraper's resonances. To account for this, we sum the synthetic impulse response calculated using morphed parameters with the measured impulse response of the scraping object weighted by $\eta$, which determines the relative amplitude of the surface and scraper impulse responses.

Since each position on the surface has a different impulse response, and because the scraper moves continuously, the traditional linear view of the sound waveform being the result of a convolution of force and a fixed impulse response does not hold. To accommodate a different impulse response at each location, we calculate the waveform $y(t)$ using a modified convolution:
\begin{align}
h^{\text{sur}}_{x,y}(\tau) &= \begin{cases}
\sum\limits_{i=1}^{50} A_i(\tau) \sin(2\pi f_{i} \tau) & 0 \leq  \tau \leq t_0 \\
0 & otherwise
\end{cases} \\
h^{\text{tot}}_{x,y}(\tau) &= h^{\text{sur}}_{x,y}(\tau) + h^{\text{scr}}(\tau)\\
f(t) &= \begin{cases}
f_{tot}(t) & 0\leq  t \leq t_1 \\
0 & otherwise
\end{cases} \\
 y(t) &= \begin{cases}
\sum\limits_{m=1}^{t} h^{\text{tot}}_{x,y}(t-m)f(m)  & 0 \leq  t\leq t_0 \\
\sum\limits_{m=1}^{t_0} h^{\text{tot}}_{x,y}(t_0-m)f(t+m-t_0) & t_0\leq t\leq t_1  \\
\sum\limits_{m=1}^{t_2-t} h^{\text{tot}}_{x,y}(t_0-m)f(t+m-t_0) & t_1\leq t \leq t_2
\end{cases}
\end{align}
where $t_2 = t_0+t_1$, $h^{\text{scr}}(t)$ has the same form as $h^{\text{sur}}_{x,y}(t)$ but is independent of position, $h^{\text{tot}}_{x,y}(t)$ is the total impulse response when the scraper is at $(x,y)$, and $f_{\text{total}}$ is the excitation at the surface.

\section{Perception of Synthetic Scraping}
\label{section:perceptscrape}

To assess whether our synthesis model produces perceptually compelling scraping sounds, we conducted two psychophysical experiments in which listeners made judgments about synthesized scraping sounds. We compared our synthesis model to several `ablated' versions of the model, in which we omitted  model components in order to test their perceptual relevance. For baselines, we also compared our model to our implementations of previous synthesis methods based on (1) physical synthesis \cite{traer2019contact}, (2) signal manipulation \cite{conan2014intuitive}, and (3) minimalist modulated noise that has previously been shown to elicit accurate motion judgments \cite{thoret2014soundtoshape}. These experiments were conducted online using Amazon's Mechanical Turk platform, using a standardized test to verify that participants were wearing headphones \cite{woods2017headphone}. We have previously found that online participants can perform about as well as in-laboratory participants \cite{woods2018,mcwalter2019,mcpherson2020} provided basic steps are taken to maximize the chances of reasonable sound presentation by testing for earphone/headphone use and to ensure compliance with instructions. All participants had self-reported normal hearing and had IP addresses from North America.

\subsection{Experiment 1. Realism of synthetic scraping sounds}

\begin{figure}[ht]
\centerline{\includegraphics[scale=0.6]{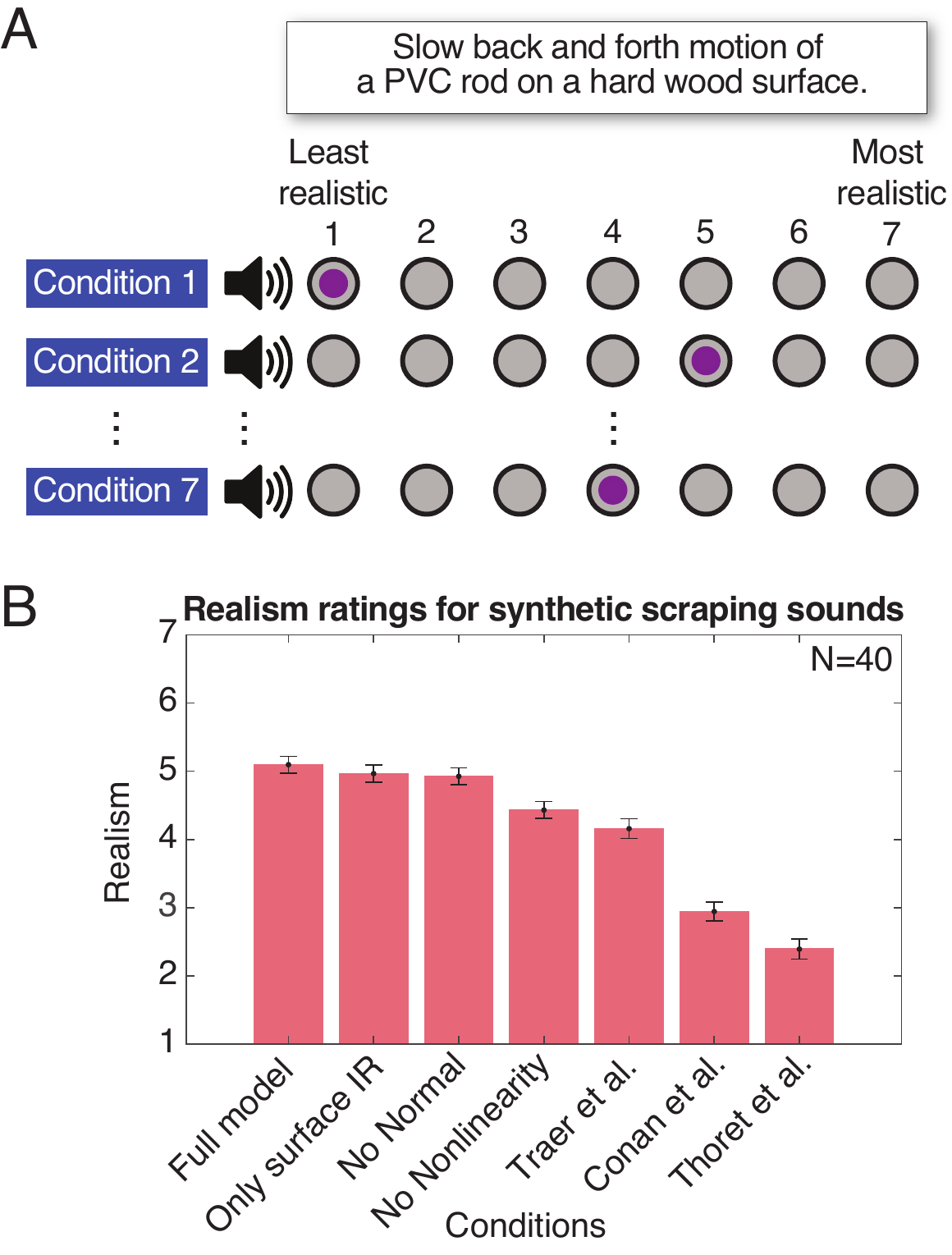}}
\caption{\label{fig:Exp1_res}{\it Experiment 1: Realism of synthesized scraping. A) Participants (N=40; mean age = 38.3 years) rated the realism of 7 different renderings of an object scraped back-and-forth over a surface, using a MUSHRA paradigm. Each rendering was via a different synthesis method. Participants were given a text description of the scraping event. B) Results of Experiment 1, showing mean realism for each synthesis method. Error bars plot SEM.}}
\end{figure}

We first tested whether our model could convincingly render descriptions of physical scraping events, by asking listeners to rate the realism of synthetic scrapes. Participants were presented with a text description of the scraping event, which specified the material of the scraper, its motion, and the material of the surface (Fig.~\ref{fig:Exp1_res}A, top). The scraper material was either plastic (PVC) or poplar wood and the surface material was basswood, poplar wood, or ceramic. The motion was one of five variants: slow back-and-forth, fast back-and-forth, circular, short single scrape, and long single scrape. Each listener evaluated a total of 30 trials comprising a fully-crossed set of these parameters (2 scraper materials $\times$ 3 surface materials $\times$ 5 motions). For each description, listeners rated the realism of seven synthetic sounds, each generated with a different method. The  methods were (1) the full model, (2) an ablated model using only the surface impulse response ($\eta=0$), (3) an ablated model lacking variation in the normal force (omitting \S \ref{section:nforce}) (4) an ablated model without a non-linearity to constrain the trajectory curvatures (omitting \S \ref{section:nonlinearity}), and (5-7) the baseline models. All seven sounds were presented in a single trial using a MUSHRA-like paradigm (Fig.~\ref{fig:Exp1_res}A). The order of the sounds was randomized on each trial. We did not include audio recordings in this experiment because they typically have distinguishing features that do not relate to scraping (e.g. room noise).

The results (Fig. \ref{fig:Exp1_res}B) show that our model produces scraping sounds that are more realistic than previous scraping synthesis methods. The non-linearity ablation caused the biggest decrease in the realism. The `Only surface IR' and normal force ablations produced relatively small impairments of the realism, potentially because there exist alternative realistic physical explanations of those sounds (e.g. damping of the scraper).%The backscraping and normal force lesions produced very subtle decreases as a result of their dependence on the material and motion conditions.

\subsection{Experiment 2. Perception of motion from scraping sounds}

We next asked whether participants could infer the motion that produced the scrape. In addition to testing whether the synthesis conveys an important physical parameter, this experiment provides a more fine-grained examination of the effect of the various ablations as it allows analysis of confusions between different motions. 

\begin{figure*}[ht]
\center
\includegraphics[width=5.3in]{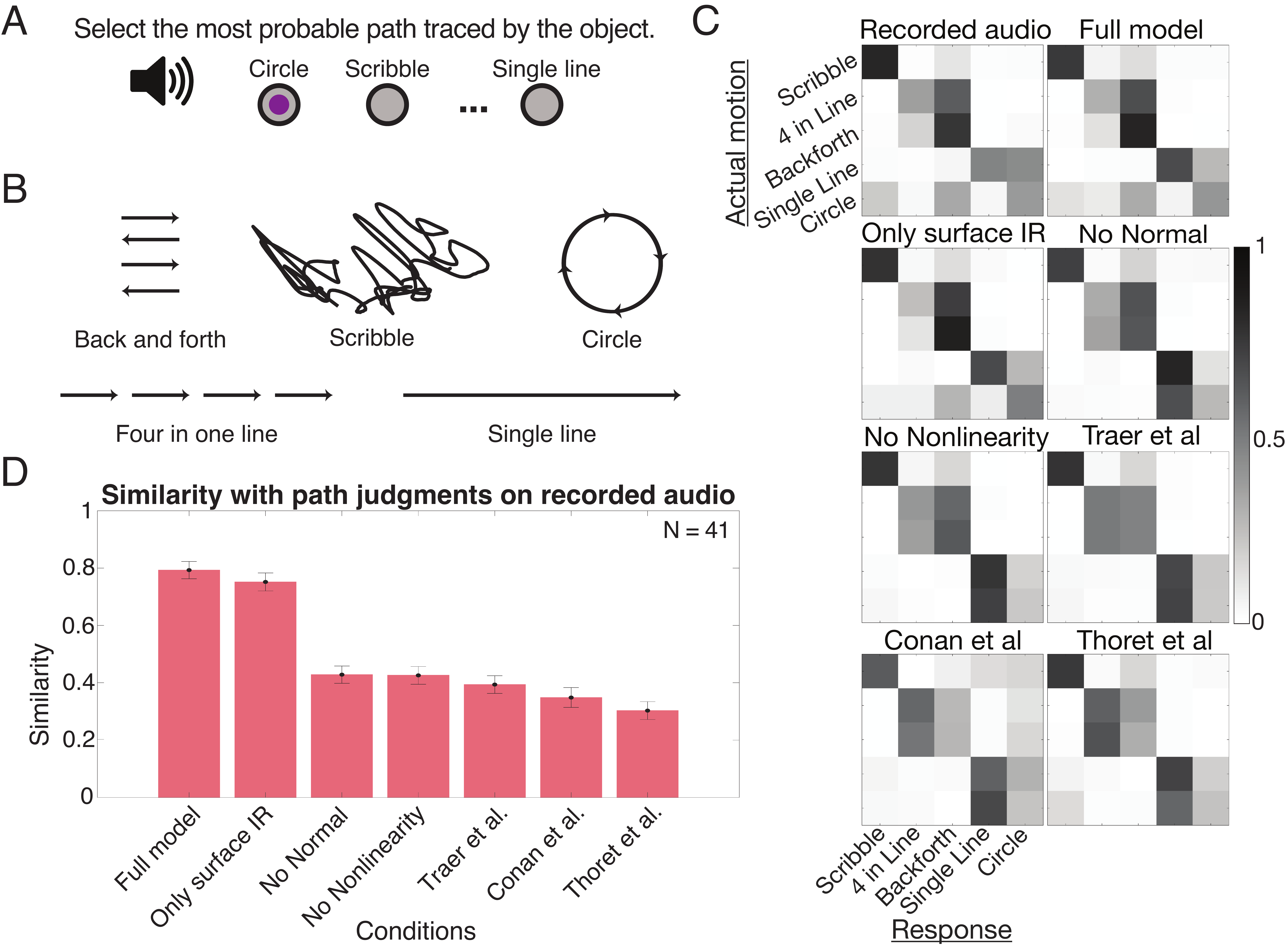}
\caption{\label{fig:expt2}{\it Experiment 2: Motion recognition from scraping. A) Participants (N=41; mean age = 40.3 years) listened to a sound and selected the path traced by the scraping object. B) Sounds were generated for each of five possible motions. C) Confusion matrices for each of the eight conditions. D) Similarity of the confusion matrix of each synthesized condition compared to that for the recorded sounds.}}
\end{figure*}

Listeners were presented with a single scraping sound and then asked to choose the most probable path (Fig. \ref{fig:expt2}A) traced by the object from five options: 1) a scribble, 2) four scrapes in a straight line, 3) four back-and-forth scrapes, 4) a single long line, or 5) a circle (Fig. \ref{fig:expt2}B). The scribble motion was obtained by optical tracking \cite{traer2019contact} and the other motions were modelled using ideal trajectories. We chose these motions because they provide pairs which share a velocity profile, but differ spatially: back-and-forth and four-in-one line, and the circle and single line. Based on previous literature \cite{thoret2014soundtoshape} we expected that listeners would make mistakes within these pairs but would make accurate judgments between them. We tested the seven synthesis conditions of Experiment 1 as well as real audio recordings. 
%Is there anything else we want to say about why these motions were chosen? Impulse response spatial distribution?

As expected, the confusion matrices (Fig. \ref{fig:expt2}C) show that listeners experienced confusions when listening to recorded as well as synthesized audio. However, the confusion patterns for our full model are more similar to those of recorded audio than were the other synthesis methods. We quantified this similarity between the confusion matrix for recorded audio $C_{\text{recorded}}$ and  each synthesis method $C_{\text{synth,i}}$ as follows: 

\begin{align}
     Similarity(i) = 1 - \frac{\|C_{\text{recorded}} - C_{\text{synth,i}}\|_2}{\|C_{\text{recorded}}\|_2}
\end{align}

As shown in Figure \ref{fig:expt2}D, the full model achieves the highest similarity to listeners' judgments on recorded audio. The `Only surface IR' model performs similarly to the full model, while the normal force and non-linearity ablations greatly decrease the similarity. These results indicate that capturing naturalistic variation in the normal force and constraining the trajectory curvature are important for the perceptual estimation of motion from scraping.

\section{Rolling Sound Synthesis Model}
\label{section:rolling}
\begin{figure}[ht]
\centerline{\includegraphics[scale=0.8]{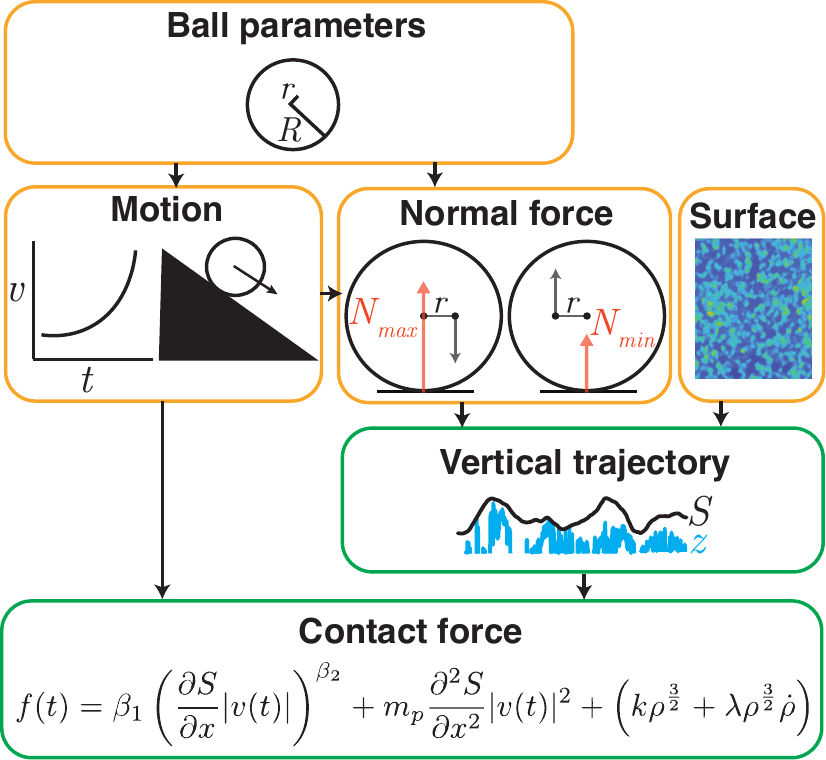}}
\caption{\label{fig:rollingforce}{\it Synthesis of contact force for rolling. Yellow boxes are inputs to the synthesizer, which are combined to yield the contact force. The contact force is combined with location-specific impulse responses in the same manner as for scraping (shown in Figure 1).}}
\end{figure}

To synthesize rolling sounds, we use a similar source-filter model where the filter changes with the change in the position of the rolling object, but with a contact force that differs from that for scraping. Pure rolling is a unique form of sustained contact where the surface is excited without the point of contact on the object being in relative motion with the surface. Previous efforts to synthesize rolling sounds have proposed a force term arising from the offset of the object's centre of mass from its geometric center (due to slight deviations from perfect sphericity). We hypothesized that there would be an additional contribution from a scraping-like term due to the catching and release between surface asperities that would depend on the trajectory of the point in contact. Figure \ref{fig:rollingforce} depicts a schematic of the overall contact force for rolling. The first two terms reflect the same calculation used in scraping, and the third term is unique to rolling.
 
\subsection{Contact force for rolling}
We used a rolling-specific force term similar to that previously proposed \cite{rath2005continuous,van2001foley}, including a non-linear dissipative component. Our implementation of this force differed from prior work in being based on the trajectory of the ball, derived as in our scraping model as a nonlinear function of the surface depth map:
\begin{align}
  f_{\text{roll}}(\rho,\dot \rho) &= k\rho^{3/2} + \lambda \rho^{3/2} \dot \rho
  \end{align}
  where $\rho$ denotes penetration depth (the deviation of the ball center of mass from its mean vertical position), $k$ denotes the equivalent spring constant of the ball material and $\lambda$ is a dissipation constant which we set by ear (to 0.1). $\rho$ and its derivative are given by
  \begin{align}
  \rho &= R - r \cos{\frac{x}{R}} +S(x,y) \\
  \dot \rho &= \frac{r}{R} \dot x \sin{\frac{x}{R}} + \dot x \frac{\partial S(x,y)}{\partial x}
\end{align}
where $x$ is the horizontal position of the center of mass:
  \begin{align}
  x &= R\theta - r\sin \theta 
\end{align}
with $r$ the distance between the center of mass and the geometric center, $R$ the mean radius of the ball, and $\theta$ the angular displacement of the ball.

The total contact force is given by
\begin{align}
f(t) = \beta_1 \left|\frac{\partial S}{\partial x} v(t)\right|^{\beta_2} +m_p\frac{\partial^2 S}{\partial x^2}|v(t)|^2 + (k\rho^{3/2} + \lambda \rho^{3/2} \dot \rho) 
\end{align}
The relative contribution of the scraping and rolling terms likely depends on how much slip is present in the rolling motion and the roughness of the materials, and can be adjusted using the existing free parameters $\beta_1$, $\beta_2$, $k$ and $\lambda$ . 

For all three force terms, the vertical trajectory $S(x,y)$ of the point of contact on the surface is determined as in the scraping model. A tanh non-linearity was used to constrain the curvatures of the vertical trajectory based on the amount of normal force. The only difference is that the normal force varies periodically over the rolling interaction due to the offset between the center of mass and the geometric center of the sphere, being maximal when the centre of mass moves downwards, and minimal when it moves up (Figure \ref{fig:rollingforce}). This variation in the normal force affects the penetration of the ball into the surface, which we modeled by varying the non-linearity parameter $\alpha$ and the subsequent smoothing.

\subsection{Impulse responses}
Location-dependent impulse responses for rolling were synthesized in the same way as for scraping. An impulse response for the rolling object is added to the surface impulse response, with the relative weighting depending on the ball material. At present we set the relative weighting by hand. In the rolling sound recordings we sought to emulate in our experiments, the surfaces were planks, and typically had a much higher contribution to the overall sound than the balls. If modeling the sound of a ball rolling on a floor, which is typically damped, the weighting would instead upweight the contribution of the ball.

\section{Perception of synthetic rolling}
\label{section:rollpercept}

To evaluate the synthesis model, we asked human listeners to rate the realism of synthetic sounds generated in various ways, using the MUSHRA-like paradigm from Experiment 1. We again compared our model to ablated models with omitted components and to two previous baselines using types of physics-based \cite{rath2005continuous} and signal-based synthesis \cite{conan2014intuitive}.

\begin{figure}[ht]
\centerline{\includegraphics[scale=0.55]{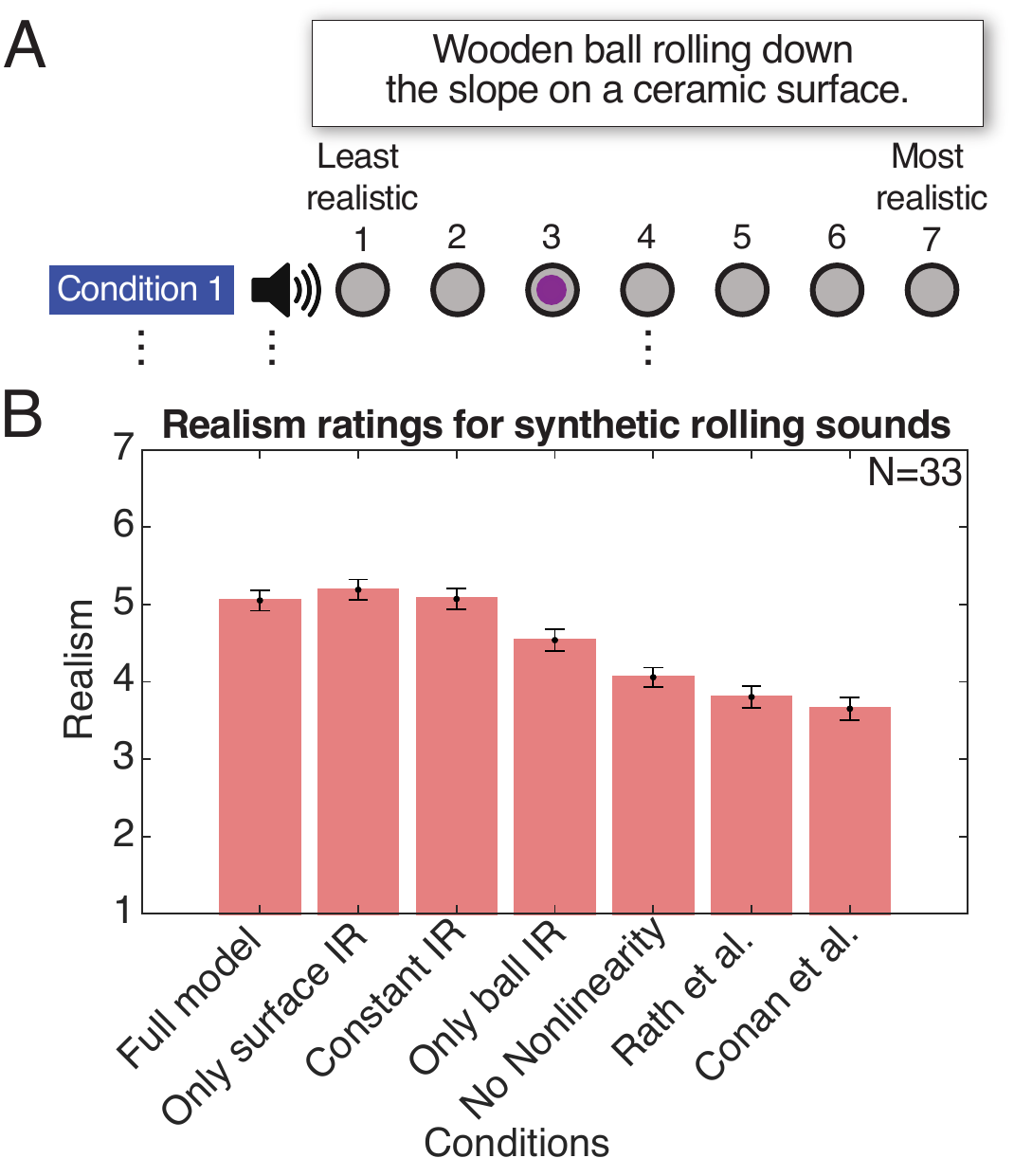}}
\caption{\label{fig:Exp3_res}{\it Experiment 3: Realism of synthesized rolling. A) Participants (N=33; mean age = 36.2 years) rated the realism of 7 different renderings of a ball rolling down or up an inclined surface, using a MUSHRA paradigm. Each of the 7 renderings was from a different synthesis method. The ball and surface varied in material. B) Results of Experiment 3, showing mean realism for each synthesis method. Error bars plot SEM.}}
\end{figure}

\subsection{Experiment 3. Realism of synthetic rolling sounds}

In each trial, online participants were presented with a test description of a rolling event which specified the material of the rolling object and the surface, incline of the surface and the motion of the object (Figure \ref{fig:Exp3_res}A). The ball material was ceramic, glass or wood. The surface material was ceramic, bass wood or poplar wood. The surface had a gradual incline in all the sounds and the ball was rolled up or down the incline. The different synthesis conditions were (1) full model, ablated models with (2) only the surface IR contribution, (3) a constant surface IR (omitting the location-dependent morphing), (4) only the ball IR contribution, (5) no nonlinearity in the calculation of contact force from the depth map, and (6-7) baseline methods \cite{rath2005continuous,conan2014intuitive}. 

As shown in Figure \ref{fig:Exp3_res}B, our model produced more realistic rolling sounds than the previous baseline models. The surface resonance was more important than the ball resonance for overall realism (`Only Surface IR' vs. `Only Ball IR'). As with the scraping sounds of Experiment 1, the non-linearity had a significant impact on the realism of rolling sounds (without it, sounds were unrealistically rough). In contrast, morphing the impulse responses did not significantly contribute to the realism.

\section{Discussion}

We developed a novel method for synthesizing scraping and rolling sounds and evaluated it with perceptual experiments. The model is object-based, in the sense of depending on relatively macroscopic properties of objects and their motions. The key innovations compared to previous methods are the introduction of a nonlinearity in the relationship between contact force and the surface depth map and the use of normal forces that more accurately reflect common scraping motions. We combine these two ideas with variants of several previous proposals for synthesis in this domain. The resulting synthesis is substantially more realistic than previous methods.

Our experiments explored the contribution of different components of the synthesis model. We found that the non-linearity used to constrain the curvatures of the vertical trajectory of the scraper (and thus the resulting contact force) was critical. Without the nonlinearity, the contact force is unrealistically high and the sounds are unrealistically rough. 

The normal force variation was also important for compelling synthesis. Its effect was most evident in the motion estimation experiment, where it was necessary to convey motion comparable to actual audio recordings. Motion recognition for the full model containing this normal force variation qualitatively matched that for audio recordings, despite using ideal velocity profiles which are not completely faithful to human-generated profiles \cite{thoret2014soundtoshape}.

Several synthesis components did not produce a clear benefit in some of the experiments, but this may be be a limitation of the experiments rather than of the synthesis. For instance, realism judgments for scraping did not distinguish between the full model and the ablated model without the normal force variation. This is plausibly because sounds without variation in the normal force correspond to an alternative physically possible situation in which the normal force does not change over the path. Participants might rate the sound as realistic because they can envision a physical scenario that could have produced the sound, even though that scenario deviates from the one that we intended. The same issue may explain why there was little effect of omitting the contribution of the scraper impulse response, as this could correspond to a situation with damping.  

We suspect that the failure to demonstrate a benefit of location-dependent impulse response morphing (Experiment 3; Fig. \ref{fig:Exp3_res}) is also a limitation of the realism judgments we used. There are realistic physical explanations for both the cases where impulse responses change with position and where they do not. For example, rolling on a damped floor would not lead to location-dependent impulse responses, whereas rolling on a small wooden plank yields a much more appreciable location-dependence. Overall, our subjective sense is that all components of the synthesis that we tested here contribute meaningfully to the resulting sound, but additional work is needed to better understand and demonstrate the situations in which each component of the synthesis is most perceptually relevant. 

The model presented here can be further improved in several respects. At present several parameters of the model are set by hand. For instance, we currently lack an empirical relationship between the non-linearity parameter and the normal force. We plan to investigate the impact of the softness/hardness of the materials empirically and to base this parameter on empirical measurements in the future. At present we also set the balance between the rolling and scraping components of rolling sounds by hand, when this should ultimately be determined by physical parameters of the motion and surface characteristics. It also seems likely that the perception of scraping and rolling is based on summary statistics \cite{ mcdermott2013summary} that capture surface roughness (as opposed to detailed representations of the underlying depth maps). Synthesis could thus plausibly be based on depth maps synthesized from summary statistics, which would make for a more parsimonious model.

Efficient and high-fidelity synthesis will open the door to new perceptual studies of ecological audition by enabling sound generation for physical events whose properties can be varied in a controlled manner. Relatively simple synthesis methods with input variables that are both physically meaningful and perceptually consequential can also provide the foundation for models of human auditory perception. If prior distributions are determined for the physical variables, the principles of Bayesian inference can be leveraged to invert the synthesis method and thereby estimate latent physical variables from sound \cite{cusimano2018auditory,kersten2003inference}. We plan to develop a complete contact sound synthesis model (including scraping, rolling, and impacts) within an inference framework that can be used to make predictions about the mechanisms underlying human physical inference. In principle, this could be combined with inverse graphics \cite{yildirim2020efficient,gan2020threedworld} to achieve multimodal physical event perception. Bayesian inference could also provide a way to set the values of the free parameters, given corpora of recorded contact sounds.
%\section{Conclusion}
%We have presented a method to synthesize the sounds of scraping and rolling from physical descriptions of objects (surface depth maps, object impulse responses) and their motions. The method is relatively efficient and produces substantially more realistic scraping and rolling sounds than previous methods.
%\nocite{*}
\bibliographystyle{IEEEbib}
\bibliography{DAFx20_tmpl} % requires file DAFx20_tmpl.bib

\end{document}